\documentclass[aip,prl,amsmath,amssymb,amstext,reprint,citeautoscript,punctuation]{revtex4-1}
\usepackage{{graphicx,xcolor,charter, gensymb, float}}

\begin{document}

\sloppy

\title{Untangling the structural, magnetic dipole, and charge multipolar orders in Ba$_2$MgReO$_6$}

\author{Aria Mansouri Tehrani}
\affiliation{Materials Theory, ETH Zurich, Wolfgang-Pauli-Strasse 27, 8093 Zürich, Switzerland}
\author{Nicola A. Spaldin}
\email{nicola.spaldin@mat.ethz.ch}
\affiliation{Materials Theory, ETH Zurich, Wolfgang-Pauli-Strasse 27, 8093 Zürich, Switzerland}

%\date{\today}

\begin{abstract}
We present a density functional theory study of the low-temperature structural, magnetic, and proposed charge-quadrupolar ordering in the double perovskite, Ba$_2$MgReO$_6$. Ba$_2$MgReO$_6$ is a spin-orbit-driven Mott insulator with a symmetry-lowering structural phase transition at 33\,K and a canted antiferromagnetic ordering of $5d^1$ Re magnetic moments at 18\,K. Our calculations confirm the existence of the proposed charge quadrupolar order and discover an additional, previously hidden, ordered charge quadrupolar component. By separately isolating the structural distortions and the orientations of the magnetic dipoles, we determine the relationship between the charge quadrupolar, structural and magnetic orders, finding that either a local structural distortion or a specific magnetic dipole orientation is required to lower the symmetry and enable the existence of charge quadrupoles.  Our work establishes the crystal structure -- magnetic dipole -- charge multipole relationship in Ba$_2$MgReO$_6$ and related 5$d^1$ double perovskites, and illustrates a method for separating and analyzing the contributions and interactions of structural, magnetic, and charge orders beyond the usual dipole level.

\end{abstract}

%\pacs{Valid PACS appear here}% 
\maketitle 

\section{Introduction}

The ordering of structural distortions, magnetic dipole moments, atomic orbital occupancies, and electronic charges is widespread in materials and leads to well-known behaviors such as changes in the symmetry, the onset of ferromagnetism, or metal-insulator transitions. While such orderings have been extensively investigated, there remain many materials that show intriguing changes in behavior indicating the onset of order, but for which the nature of the order is not yet known. The search to find and identify such {\it hidden orders} and understand their relationship to the resulting properties is an important challenge in materials physics, often requiring developments in the state-of-the-art for both experimental or theoretical methods\cite{aeppli_hidden_2020}. 

While historically, the actinides, with their large spin-orbit coupling (SOC) and strongly magnetic $5f$ electrons, have been a rich area of exploration for hidden order\cite{magnani_inelastic_2008, suzuki_first-principles_2010, caciuffo_multipolar_2003, mydosh_hidden_2020, kusunose_hidden_2011, lander_fifty_2020, kubo_multipole_2006}, the 5$d$ transition metal oxides have recently attracted attention, since their SOC, crystal-field splitting, orbital bandwidth and electron-electron repulsion, $U$, tend to have similar energies\cite{wang_noncollinear_2017, lu_magnetism_2017}. 
Among the $5d$ compounds, the double-perovskite oxides, $A_2BB^{'}O_6$, with $B$ a non-magnetic cation and $B'$ a $5d^1$ ion are of particular interest. First, the wide separation between adjacent $B'$ cations means that they tend to have a small bandwidth leading to localized correlated electrons and insulating behavior. Second, since the $B'$ sites form a face-centered cubic lattice, their magnetic interactions can be geometrically frustrated. Indeed, a huge range of magnetic ground states have been reported in addition to conventional long-range antiferro- or ferromagnetic order, including short-range antiferromagnetic order, weak ferromagnetism, and gapped spin-singlets; for a review see Ref.~\onlinecite{marjerrison_cubic_2016}. Finally, a formally $d^1$ ion in an octahedral crystal field has exact cancellation between its effective orbital angular momentum, $l_{eff} = -1$, and spin angular momentum, $s=+1$, components\cite{khomskii_transition_2014, abragam_electron_2012}. This provides the intriguing possibility of studying the physics of higher-order magnetic multipolar states in the absence of a magnetic dipole. Considerable recent interest has been spawned by a phenomenological study of a model Hamiltonian appropriate for such $5d^1$ oxides, in which several exotic phases have been identified \cite{chen_exotic_2010}. The model contains a nearest-neighbor antiferromagnetic superexchange, $J$, a nearest-neighbor ferromagnetic exchange, $J'$, a quadrupole-quadrupole interaction, $V$ and spin-orbit coupling, $\lambda$. The authors identify, at the mean-field level, a ferromagnetic phase, an unusual antiferromagnet in which magnetic octupolar order dominates over dipolar order, and a charge quadrupolar-ordered paramagnet, as a function of these various parameters. In addition, a possible quantum-spin-liquid state is proposed when quantum fluctuations are taken into account. Experimental examples include Ba$_2$NaOsO$_6$, which exhibits magnetic octupolar interactions on its formally $5d^1$ Os$^{7+}$ ion\cite{lu_extraordinary_2017}, and Ba$_2$CdReO$_6$, in which a structural transition observed using X-ray diffraction has led to the proposal of charge quadrupolar ordering\cite{hirai_possible_2021}. 

Our focus in this work is Ba$_2$MgReO$_6$, which has been the subject of several recent studies\cite{marjerrison_cubic_2016, hirai_successive_2019, hirai_detection_2020}. Here the Mg$^{2+}$ ion is non-magnetic, and the formally Re$^{6+}$ has the $5d^1$ configuration.
At room temperature, Ba$_2$MgReO$_6$ adopts the ideal double perovskite structure with space group $Fm\bar{3}m$\cite{Sleight/Longo/Ward:1962,Bramnik_et_al:2003}. As the temperature is lowered, it is reported to undergo two successive symmetry-lowering phase transitions\cite{hirai_successive_2019}. An anomaly in heat capacity measurements at 33\,K led to a proposal of charge quadrupole ordering, although no structural distortion was initially identified\cite{hirai_successive_2019}. Later, synchrotron X-ray measurements on high-quality single crystals detected a small tetragonal structural distortion at the same temperature, pointing to the coupling of charge quadrupole ordering to the lattice\cite{hirai_detection_2020}. At lower temperature, 18\,K, there is an unusual antiferromagnetic ordering of the magnetic dipoles with a large canting estimated to be $\sim$ 40$\degree$ away from the [110] easy axis leading to a saturation moment of $\sim$ 0.3 $\mu_B$ per formula unit \cite{hirai_successive_2019, lu_magnetism_2017}. Ref.~\onlinecite{hirai_detection_2020} proposed a possible relationship between the unusual magnetic order and the quadrupolar order which is already established above the N\'eel temperature.
 
Here, we present a detailed density functional theory (DFT) study of the structural, electronic, and magnetic properties of Ba$_2$ReMgO$_6$. Our primary goal is to provide an understanding of the nature and interplay between the structural distortions, the charge quadrupole ordering, and the magnetic dipole ordering, as well as to determine the driving force for the occurrence of the charge quadrupole ordering. We achieve this goal by calculating the magnitude and sign of the various quadrupole components, with and without crystallographic distortions, and in the ordered and disordered magnetic states. This in turn allows us to analyze the dependence of the charge quadrupoles in Ba$_2$ReMgO$_6$ on the magnetic order and the symmetry-lowering crystallographic distortions separately. Importantly, DFT allows us to accurately describe the detailed rhenium-oxygen chemistry, which dictates these complex phenomena in Ba$_2$MgReO$_6$.

A second goal is methodological and motivated by the fact that the charge quadrupole ordering occurs at a higher temperature than the magnetic dipolar ordering in Ba$_2$ReMgO$_6$. Therefore an understanding of the interplay between the multipolar ordering, the crystallographic distortions, and the electronic structure is most appropriately achieved by studying the paramagnetic state. Conventional DFT approaches, however, apply periodic boundary conditions to a single unit cell and do not incorporate temperature effects, and therefore are not suitable for describing paramagnetic local-moment insulators. Here we explore the use of a supercell approach that has recently been used successfully to obtain paramagnetic and insulating behavior in materials that would be metallic within the standard DFT framework~\cite{yoon_non-dynamical_2019, varignon_mott_2019, zhang_symmetry-breaking_2020}. We find that the approach is also useful here, allowing us to generate paramagnetic configurations with disordered Re local moments and in turn to study the formation and ordering of quadrupoles in the magnetically disordered state.

\section{Computational details}

All DFT calculations were performed using the Vienna $ab-initio$ simulation package (VASP) based on a plane-wave basis set and projector augmented wave (PAW) pseudopotentials\cite{kresse_ab_1993, kresse_efficient_1996, kresse_ultrasoft_1999, blochl_projector_1994}. The following electrons were included as valence states: Ba: $5s^2\,5p^6\,6s^2$, Mg: $2p^6\,3s^2$, Re: $5p^6\,5d^5\,6s^2$, and O: $2s^2\,2p^4$. To approximate the exchange and correlation, the Perdew-Burke-Ernzerhof (PBE) implementation of the generalized gradient approximation was used\cite{perdew_generalized_1996} with an on-site effective Hubbard $U_{eff}$ = $U\,-\,J$ correction for the Re $d$ orbitals within the Dudarev approach\cite{dudarev_electron-energy-loss_1998}. Spin-orbit coupling was included within the fully relativistic scheme in which $H_{SOC} = \sum_{i} \xi_i\boldsymbol{\mathbf{l}}_i.\boldsymbol{\mathbf{s}}_i$, coupling the spin ($\boldsymbol{\mathbf{s}}_i$) and angular momentum ($\boldsymbol{\mathbf{l}}_i$) operators is added to the Hamiltonian; the $\xi_i$ are calculated from the radial derivatives of the PAW potentials\cite{steiner_calculation_2016, blanco-rey_validity_2019} and $i$ refers to the electrons in the system. Additionally, we analyzed the influence of SOC strength by linearly scaling the $\xi_i$ factors in the SOC Hamiltonian. An energy cutoff of 600\,eV and a $k$-point mesh of 6$\times$6$\times$6 was used with convergence criteria of 1\,$\times\,10^{-6}$\,eV for both electronic and ionic parts. The canting angles were determined utilizing constrained magnetic noncollinear calculations to calculate the total energy as a function of moment orientation. To fix the magnetic direction, we added a parameterized ($\lambda$) penalty to the total energy. We found that a value of $\lambda$ = 10 fixed the moments in the desired direction while introducing a negligible energy penalty of around $10^{-5}$\,eV. 

\begin{figure}
\includegraphics[width=3in]{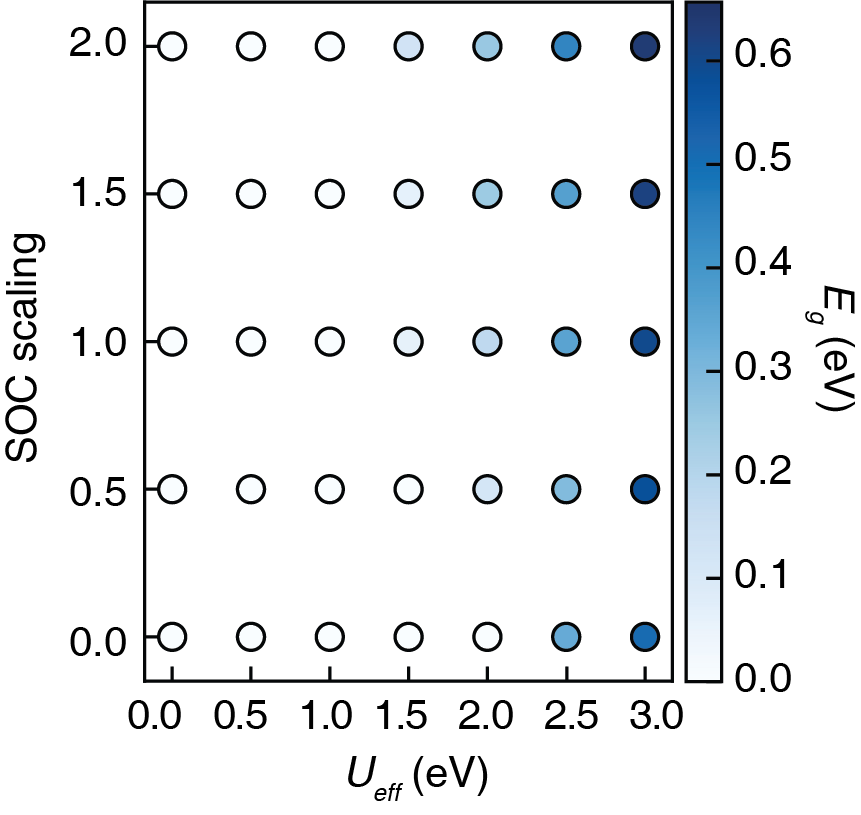}
\centering
\caption{Calculated band gap values of Ba$_2$ReMgO$_6$ as a function of SOC scaling (that is the amount by which we scale the true $\xi_i$ values) and $U_{eff}$. }
\label{fig0}
\end{figure}

Fig.\,\ref{fig0} shows the calculated band gap of Ba$_2$ReMgO$_6$, $E_g$, as a function of $U_{eff}$ and SOC strength. The blue shades in Fig.\,\ref{fig0} indicate the size of $E_g$, with  white corresponding to the metallic state and dark blue to $E_g\,\approx$\,0.6\,eV. We see that, despite the relatively large bandwidth of Re $5d$ orbitals, the large distance (4.09\,\AA) between the Re and Mg ions in the double-perovskite structure, combined with the $U_{eff}$ and the SOC leads to an insulating state for the true SOC (scaling = 1) and $U_{eff} > 1$\,eV. For most of the calculations performed we selected $U_{eff} = 1.8$\,eV and SOC scaling = 1 , where $E_g$ is calculated to be 0.2\,eV, which is very close to the 0.17\,eV thermal activation energy reported from electrical resistivity measurements\cite{hirai_successive_2019}.

The paramagnetic structures were constructed by creating 2\,$\times$\,2\,$\times$\,2 supercells containing 160 atoms with 16 Re sites. The magnetic moments on 14 Re sites were initially assigned before the remaining two were set manually to enforce the correct net magnetic moment. Similarly, in-plane paramagnetic structures were constructed with the additional constraint of only allowing $x$ and $y$ components for the magnetic vectors. Five randomized configurations (snapshots) were calculated and averaged to ensure the robustness of the data. The $k$-point mesh was reduced to 3\,$\times$\,3\,$\times$\,2 for the supercells while the same convergence parameters implemented for the unit cell were used otherwise.
Finally, the multipoles were computed by decomposition of the atomic-site density matrix ($w^{kpr}$) into irreducible spherical tensor moments using the density matrix obtained from VASP calculations. This method was initially formulated and developed for the ELK code, as described in Ref.~\onlinecite{cricchio_itinerant_2009}, and was later implemented in VASP.\cite{thole_magnetoelectric_2018} 

\section{Results and Discussions}

We begin by calculating the minimum energy zero kelvin structure and magnetic ordering within the density functional formalism.

\begin{figure}
\includegraphics[width=3in]{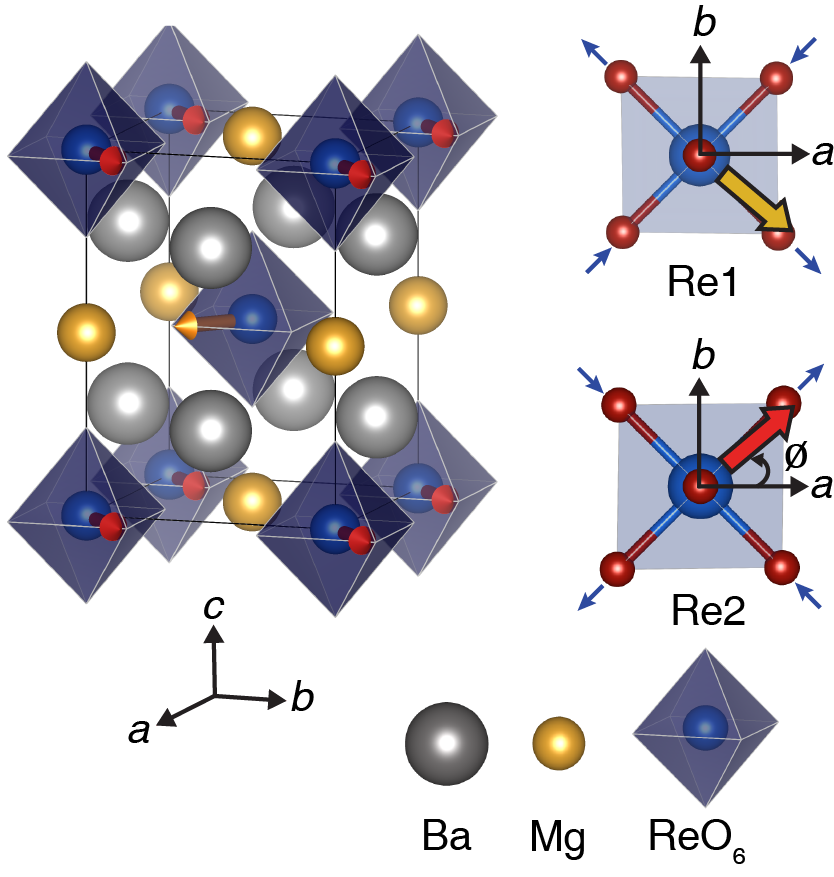}
\centering
\caption{Calculated lowest-energy structure of Ba$_2$MgReO$_6$. The space group is $P4_2/mnm$. The oxygen atoms at the corners of the ReO$_6$ octahedra are omitted for clarity. The spins are shown with yellow and red arrows for Re1 and Re2 atoms, respectively. The canting angle is indicated by $\phi$.}
\label{fig1}
\end{figure}

Full structural optimization using PBE\,+\,SOC\,+\,$U$ calculations yields the tetragonal $P4_2/mnm$ ground state structure (Fig.\,\ref{fig1}), consistent with the most recent experimental studies. Note that this is a sub-group of the cubic $Fm\bar{3}m$ structure that is reported at room temperature. The lattice parameters are calculated to be $a = 5.79$\,\AA\, and $c = 8.20$\,\AA\,, in good agreement with the experimental results obtained at 6\,K using Synchrotron X-ray single-crystal diffraction, $a = 5.70$\,\AA\, and $c = 8.09$\,\AA\cite{hirai_detection_2020}. The tetragonal structure has two inequivalent Re sites with equal but opposing distortions of their coordinating octahedra as illustrated in Fig.\,\ref{fig1}. For the rhenium site at the center of the unit cell (Re1) the Re1-O bond lengths are 1.93\,\AA\,and 1.98\,\AA\,along [110] and [$\bar{1}\bar{1}$0], respectively. In contrast, the Re2-O bonds are elongated along [110] and contracted along [$\bar{1}\bar{1}$0]. These distortions are indicated by small blue arrows in the right panels of Fig.\,\ref{fig1}. At both rhenium sites, the Re-O bond lengths along [001] are 1.96\,\AA.

Our calculated magnetic structure of Ba$_2$MgReO$_6$ is also depicted in Fig.\,\ref{fig1} with the yellow and red arrows indicating the Re1 and Re2 spin magnetic dipole moments, respectively. Our calculated magnetic ground state, in agreement with the prior experimental reports, can be described as a strongly canted antiferromagnet with a net ferromagnetic moment, or as two interpenetrating ferromagnetic sub-lattices that are non-collinear with each other so that their magnetizations partially cancel. The net ferromagnetic moment is along the tetragonal [100] direction, which corresponds to the [110] direction of the simple cubic perovskite unit cell. The computed canting angle, $\phi$, extracted by calculating the variation in total energy as we rotate the spin moments using constrained magnetic noncollinear calculations is 24\degree\,for $U_{eff} = 1.8$\,eV. This corresponds to a net magnetic moment of 0.25\,$\mu_B$ per formula unit, slightly underestimating the experimental saturation moment of 0.3\,$\mu_B$. Note that this is significantly reduced compared to the spin-only value of 1.73\,$\mu_B$ for formally $d^1$ systems due to the partial cancellation of the spin magnetic component by the unquenched orbital component\cite{hirai_successive_2019}, confirming the large unquenched orbital moment in Ba$_2$MgReO$_6$. As expected, the complete cancellation of the spin and orbital components is not realized due to the hybridization of O\,2$p$, and Re\,5$d$ orbitals.

\begin{figure}
\includegraphics[width=3in]{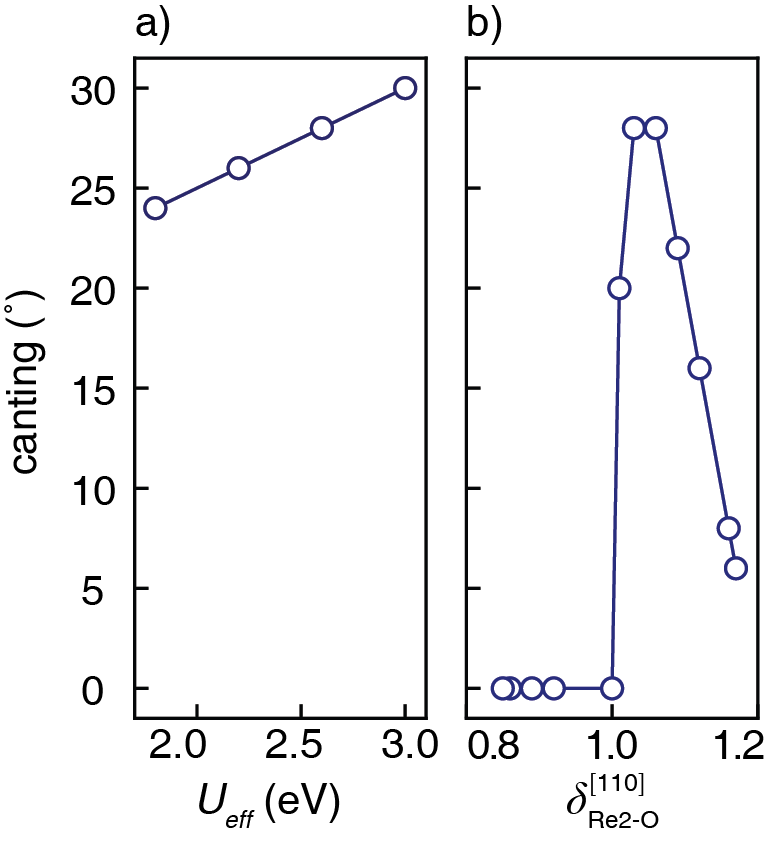}
\centering
\caption{a) Canting angle as a function of $U_{eff}$. b) Canting angle as a function  of $\delta_{\text{Re2-O}}^{[110]}$, defined as the ratio of Re2-O contraction to Re2-O expansion along [110]. $\delta_{\text{Re2-O}}^{[110]} = 1$ means all Re-O bonds are equal. distortion $\delta_{\text{Re2-O}}^{[110]} > 1$ corresponds to the cases where the in-plane Re1-O and Re2-O bonds distort in the same direction as the optimized structure. $\delta_{\text{Re2-O}}^{[110]} < 1$ occurs when the in-plane Re1-O and Re2-O bonds distort in the opposite direction from the optimized structure. In both cases the out-of-plane Re-O bonds are optimized.} 
\label{fig2}
\end{figure}

We find, however, that our calculated canting angle is dependent on our choice of $U_{eff}$, as shown in Fig.\,\ref{fig2}a. Since we expect a smaller correlation effect in 5$d$ oxides compared to the more thoroughly studied 3$d$ materials we explore $U_{eff}$ values between 1.8 and 3.0\,eV and obtain canting angles of 24 to 30 degrees, all of which slightly underestimate the estimated experimental value of 40 degrees. We find that separate treatment of the $U$ and $J$ parameters within the Liechtenstein method\cite{liechtenstein_density-functional_1995} slightly suppresses the canting angles compared with the corresponding $U_{eff}$ (for example, the canting angle with $U = 2$ and $J = 0.2$ eV is 20$\degree$, compared with 24$\degree$ for $U_{eff} = 1.8$ eV). Therefore, while our results provide the correct qualitative magnetic ground state for Ba$_2$MgReO$_6$\cite{hirai_successive_2019}, we do not make quantitative predictions of the orientation of the magnetic moments. We note also that, as the resonant elastic x-ray scattering (REXS) experiment at the Re $L_3$ edge was performed at only one azimuthal angle, the experimental estimate of 40\degree\,has a large uncertainty. An azimuthal scan or full linear polarization analysis of the scattered x-rays would be required to obtain a better refinement of the canting angle of the Re spins. 

To understand the relationship between spin canting and crystal structure distortion in Ba$_2$MgReO$_6$, we calculate the canting angle as a function of the magnitude of the oxygen octahedral distortions; our results are shown in Fig.\,\ref{fig2}b. We define a parameter, 
$$\delta_{\text{Re2-O}}^{[110]} = \frac{\text{contraction (in \AA) of Re2-O along }[110]}{\text{expansion (in \AA) of Re2-O along } [110]},$$ 
to describe the bond asymmetry arising from the structural distortion. $\delta_{\text{Re2-O}}^{[110]}$ equal to 1 in Fig.\,\ref{fig2}b corresponds to the high-temperature high-symmetry phase, in which, atomic positions are fixed to the reported high-temperature experimental coordinates\cite{hirai_detection_2020} with all Re-O bonds equivalent. We then manually vary $\delta_{\text{Re2-O}}^{[110]}$ by contracting ($\delta_{\text{Re2-O}}^{[110]}$ $< 1$) or expanding ($\delta_{\text{Re2-O}}^{[110]}$ $> 1$) the Re-O bonds in the $x-y$ plane and relaxing the Re-O bond in the $z$ direction. For the fully relaxed DFT structure, the $\delta_{\text{Re2-O}}^{[110]}$ value is 1.02. Subsequently, we determine the canting angles using the approach described previously. The canting angle is surprisingly zero not only for the undistorted configuration ($\delta_{\text{Re2-O}}^{[110]} = 1$)  but also when the Re-O bonds are distorted in the opposite direction to that found in the relaxed structure ($\delta_{\text{Re2-O}}^{[110]} < 1$). This indicates a strong coupling between the structural distortions and the canting angle that is not determined purely by the symmetry of the structure.

Our analysis of the dependence of the canting angle on the structural distortions indicates a strong coupling between the magnetism and the crystal structure in Ba$_2$MgReO$_6$. Since the charge quadrupoles also couple to the structure, we anticipate a structurally mediated coupling between magnetism and the charge quadrupoles, which we investigate next.

%charge quadrupoles

 To confirm the experimental proposal of charge quadrupole ordering and further explore the physics of this ordering we begin by calculating a quantity that represents the size and arrangement of charge quadrupoles in Ba$_2$MgReO$_6$. For a general charge density, $\rho_e$, the charge moments, $Q_{kp}$, are defined as $Q_{kp} = \sqrt{\frac{4\pi}{2l+1}}\int dr(r^lY^*_{kp}(\Theta,\Phi))\rho_e(r))$, in which the charge density, $\rho_e$, is projected onto the spherical harmonics, $Y^*_{kp}$\cite{suzuki_first-principles_2018}. Therefore, the charge quadrupoles, and other terms in the multipole expansion of the charge density, can be extracted from the electron density matrix generated in a DFT calculation by constructing the irreducible spherical tensor, $w^{kpr}$, whose components are proportional to the multipolar moment expansions of charge and magnetization densities. $w^{kpr}$ is initially constructed by considering a double tensor, $w^{kp}$, where $w^{k}$ and $w^{p}$ are the orbital multipole momentum, and the spin dependence components, respectively. However, $w^{kp}$ is neither irreducible nor includes the effect of SOC. These issues are addressed by creating the $w^{kpr}$ using the index $|k-p|<r<|k+p|$ which couples the orbital and spin components of the double tensor, hence corresponding to the total multipole moments. For example, $w^{011}$ and $w^{101}$ correspond to the spin and orbital moments, respectively. $w^{202}$ is of our particular interest as it gives the charge quadrupole moments\cite{cricchio_itinerant_2009,bultmark_multipole_2009}.

\begin{figure}
\includegraphics[width=3in]{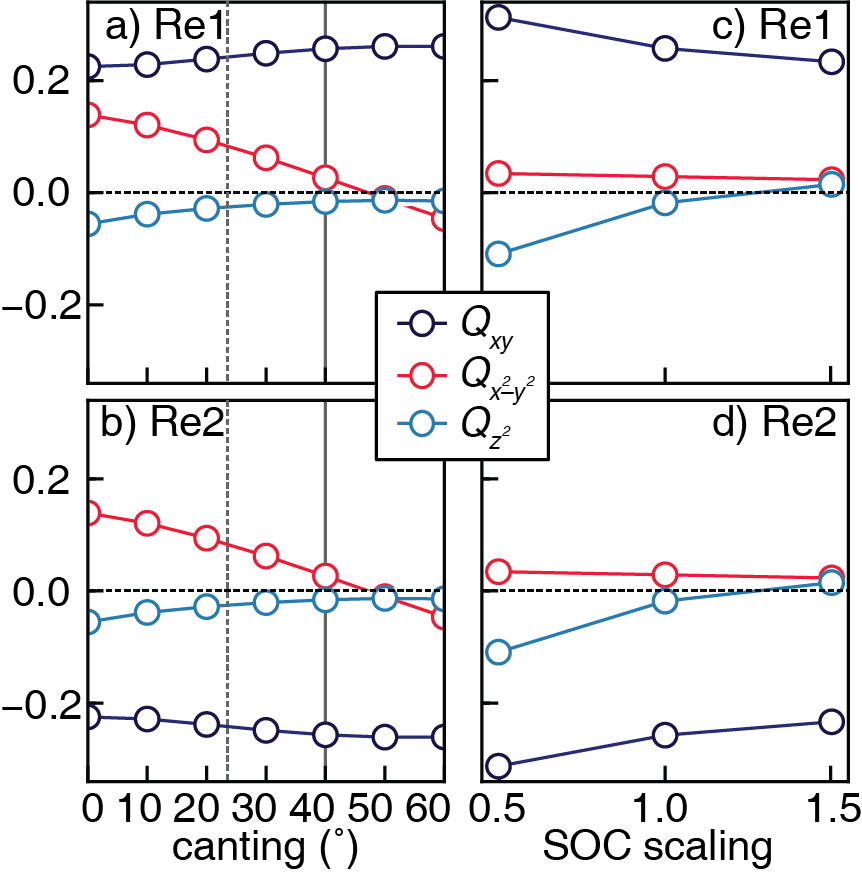}
\centering
\caption{ Expectation values of the $Q_{xy}$, $Q_{x^2 - y^2}$ and $Q_{z^2}$ charge quadrupole components in units of electron charges for Re1 (a and c) and Re2 (b and d) sites as a function of canting angle (a, b), and the scaling of SOC (c, d). The vertical dashed and solid lines in (a, b) represent our computationally determined and the previously experimentally estimated\cite{hirai_detection_2020} values of the canting angle for Ba$_2$MgReO$_6$, respectively.}
\label{fig3}
\end{figure}

First, we calculate the charge quadrupoles on the Re ions in the ground state, 0\,K crystal structure with the canted antiferromagnetism configuration described above. Our calculations reproduce the antiferroic ordering of the $Q_{xy}$ component that had previously been reported experimentally as well as the ferroic $Q_{z^2}$ order that had been inferred from the measured lattice distortions\cite{hirai_detection_2020}. Our computations reveal an additional ferroic ordering of the $Q_{x^2-y^2}$ component that had not previously been identified.  We find that $Q_{yz}$ and $Q_{xz}$ are zero, as expected by symmetry. Note that here the charge quadrupole components are described with reference to the $P4_2/mnm$ unit cell shown in Fig.\,\ref{fig1}, and our $Q_{xy}$ corresponds to the $Q_{x^2-y^2}$ of the unit cell convention used in Refs.\,\onlinecite{hirai_detection_2020, chen_exotic_2010}. We note that the general model framework for 5$d^1$ double perovskites discussed above\cite{chen_exotic_2010} only predicted the $Q_{xy}$ component. Ref\,\onlinecite{hirai_detection_2020} suggested that this could be due to their neglecting the effects of quantum fluctuations or electron-phonon couplings\cite{hirai_detection_2020}. Our DFT results demonstrate that neither quantum fluctuations nor electron-phonon coupling is required to capture the additional charge quadrupole components.

To understand how the charge quadrupoles and their ordering are affected by the orientation of the magnetic dipole moments, we next calculate the expectation values of the non-zero components of the charge quadrupoles as a function of canting angle, Fig.\,\ref{fig3}a,b. Here we see that the magnitude and sign of the $Q_{x^2-y^2}$ quadrupoles are strongly dependent on the canting angle. In contrast, $Q_{xy}$ and $Q_{z^2}$ show almost negligible change, $<0.02$\,($e$), as the canting angle increases from 0 to 60 degrees. Note that the $Q_{x^2-y^2}$ component vanishes when the canting angle is $\approx\,50^{\circ}$, (close to the experimentally estimated $40^{\circ}$) emphasizing the significance of an accurate determination of the canting angle. Next, we calculate the dependence of the charge quadrupoles on the SOC scaling. Interestingly, we find that those quadrupole components that depend strongly on the canting angle depend only weakly on the SOC strength and vice versa. As the SOC scaling is varied (we show the range in which it is reduced to and increased by one half of the full amount), $Q_{x^2-y^2}$ remains almost constant, whereas $Q_{xy}$ and $Q_{z^2}$ vary by $>0.1$\,($e$), (Fig.\,\ref{fig3}c,d). Thus, we conclude that $Q_{xy}$ and $Q_{z^2}$ are influenced by the SOC, whereas $Q_{x^2-y^2}$ is coupled to the canting angle. We note that this strong dependence of the size and even the sign of the charge quadrupoles on the details of the magnetic structure (which we modify by tuning the canting angle) and on the coupling of the spins to the lattice (which we modify by tuning the spin-orbit coupling) illustrate the sensitivity of the charge quadrupoles to properties other than just the crystal structure. Consequently, experimental probes, such as X-ray diffraction, which are sensitive primarily to the structure, might yield incomplete information regarding the presence or ordering of quadrupoles. In this context,  DFT studies can play an important role in identifying multipolar ordering in crystalline materials, and in guiding the choice of an appropriate probe for experimental verification.

\begin{figure}
\includegraphics[width=3in]{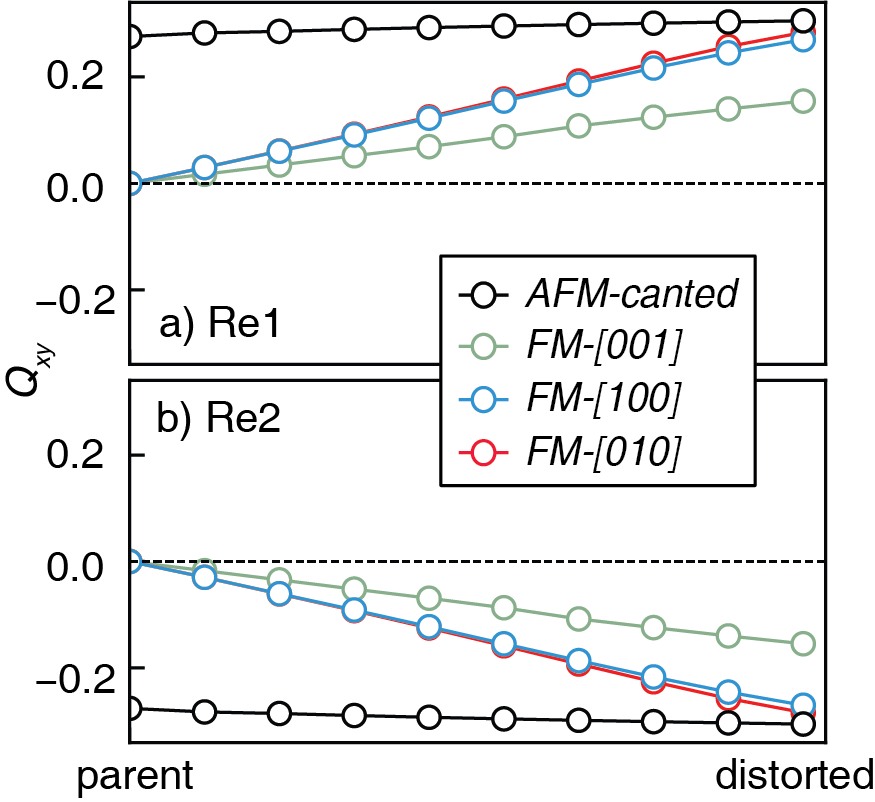}
\centering
\caption{Evolution of the $Q_{xy}$ on the Re ions as the $Fm\bar{3}m$ parent structure is distorted to the low temperature $P4_2/mnm$ structure for a) Re1 and b) Re2 atoms. Values for four different magnetic orderings canted antiferromagnetic configuration (canting angle = 40\degree) and ferromagnetic configurations along [001], [100], and [010] directions are shown. The unit of $Q_{xy}$ is in electron charges.} 
\label{fig_new}
\end{figure}

While the structural distortion and the ordering of charge quadrupoles occur at the same temperature, it is unclear whether the charge quadrupole ordering is driven by the structural distortion or vice versa. To address this question, we next calculate the evolution of the $Q_{xy}$ component of the Re charge quadrupoles as we evolve the structure from the $Fm\bar{3}m$ high-temperature experimental structure\cite{hirai_detection_2020} to the low-symmetry $P4_2/mnm$ DFT structure by linearly interpolating the atomic positions between the two phases. Our results are shown in Fig.\,\ref{fig_new} for four different magnetic orderings: The canted antiferromagnetic configuration, with canting angle = 40\degree\, and three ferromagnetic configurations with moments oriented along [001], [100], and [010] respectively. We see that the antiferroic ordering of $Q_{xy}$, which we observed previously in the $P4_2/mnm$ canted-AFM phase, is retained as the distortion is reduced, with only a small decrease in the values of the local $Q_{xy}$ in the high-symmetry parent phase. For all of the ferromagnetic cases, however, $Q_{xy}$ goes linearly to zero as the structural distortion is reduced. Therefore we conclude that either a structural distortion or a magnetic ordering that lowers the symmetry appropriately is required for the formation of charge quadrupoles. We note, however, that the charge quadrupole ordering in Ba$_2$MgReO$_6$ occurs at a higher temperature than the magnetic ordering, indicating that structural distortion plays a critical role.

\begin{figure}
\includegraphics[width=3in]{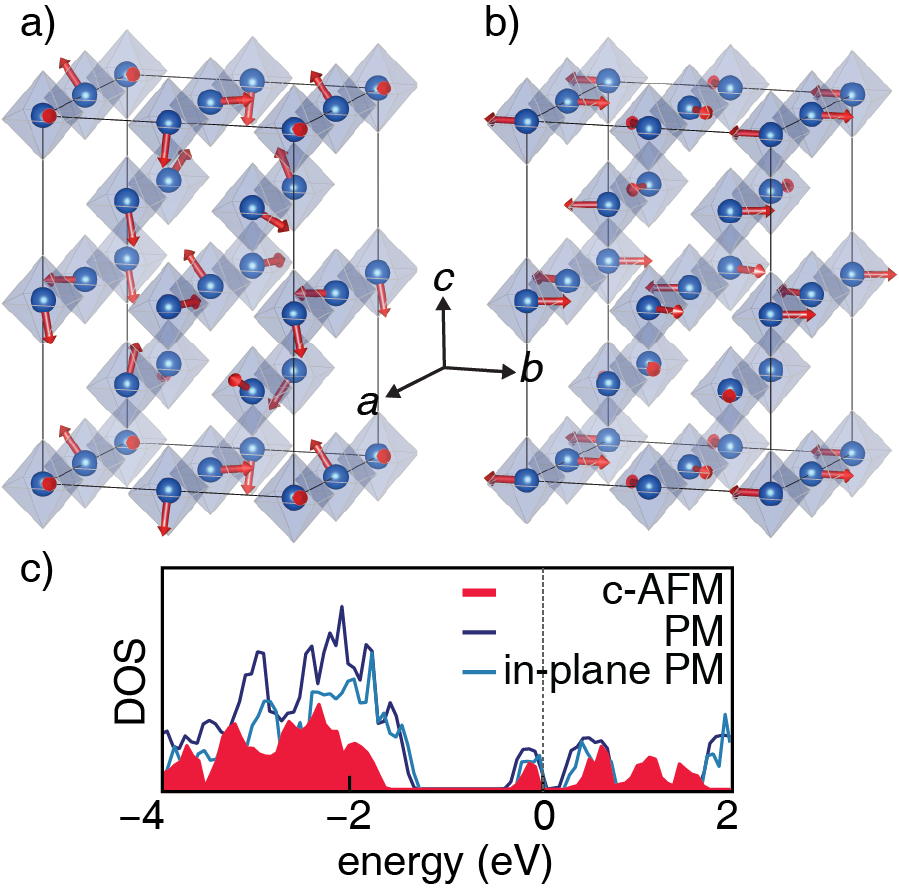}
\centering
\caption{a) Schematic of the paramagnetic supercell of Ba$_2$MgReO$_6$, constructed with the structure of the low-symmetry $P4_2/mnm$ phase. Magnetic dipole moments are shown by red arrows. Only ReO$_6$ octahedra are shown. b) Schematic of the in-plane paramagnetic supercell of Ba$_2$MgReO$_6$, again constructed with the structure of the low-symmetry $P4_2/mnm$ phase. Spins are shown by red arrows. Only ReO$_6$ octahedra are shown. c) DOS of the canted antiferromagnetic unitcell (c-AFM), the paramagnetic supercell (PM), and the in-plane paramagnetic supercell (in-plane PM), all with $U_{eff} = 1.8$\,eV. The Fermi level is indicated by the vertical dashed line at zero. Note the existence of a band gap in all three cases.} 
\label{fig4}
\end{figure}

Given that the onset of charge quadrupole ordering occurs in the paramagnetic phase, it is clear that further insight can only be gained by calculating the behavior of the charge quadrupoles in structures that do not have magnetic dipole ordering. We explore this direction next.
As expected, a standard paramagnetic DFT calculation for the unit cell of Fig.\,\ref{fig1} does not appropriately describe a disordered local-moment system and leads to a metallic state. Instead, we follow a similar approach to Refs.\cite{yoon_non-dynamical_2019, varignon_mott_2019, zhang_symmetry-breaking_2020},  and construct a 160-atom paramagnetic supercell containing randomly disordered local spin orientations, as shown in Fig.\,\ref{fig4}a, where the red arrows indicate the magnetic moments on the 16 rhenium atoms (blue circles). The orientations and sizes of the magnetic moments are constrained in our calculations so that the total magnetic moment of the supercell is zero. The ReO$_6$ polyhedra are displayed by the light blue shade while Ba, Mg, and O atoms are eliminated from the figure for clarity. Since we find that the magnetic (001) easy plane is significantly favored, by 25 meV per formula unit, over the [001] axis, we also investigate the behavior of a phase in which the spins are disordered but lie in-plane (Fig.\,\ref{fig4}b); we refer to this arrangement as in-plane paramagnetic in the following. This structure is constructed using the same constrained-spins method as the fully paramagnetic structure, but with the spins required to lie within the (001) crystallographic plane.

Importantly, for both the paramagnetic and the in-plane paramagnetic cases, the calculated density of states displays a band gap even with a relatively small $U_{eff}$ value of 1.8\,eV (Fig.\,\ref{fig4}c), indicating that this approach captures the insulating state while being paramagnetic. Fig.\,\ref{fig4}c also compares the density of states of these supercells to the magnetically ordered (canted antiferromagnetic, c-AFM) phase. The densities of states remain similar with a small shift to higher energies at $\approx1\,eV$ below the Fermi energy for the paramagnetic and the in-plane paramagnetic supercells as well as a slight decrease in the band gap, especially for the fully disordered case. The success of the disordered local-moment supercell approach in describing the Mott-insulating physics of Ba$_2$MgReO$_6$ in its paramagnetic state enables us to study the charge quadrupoles in the absence of magnetic order.

\begin{figure*}
\includegraphics[width=6in]{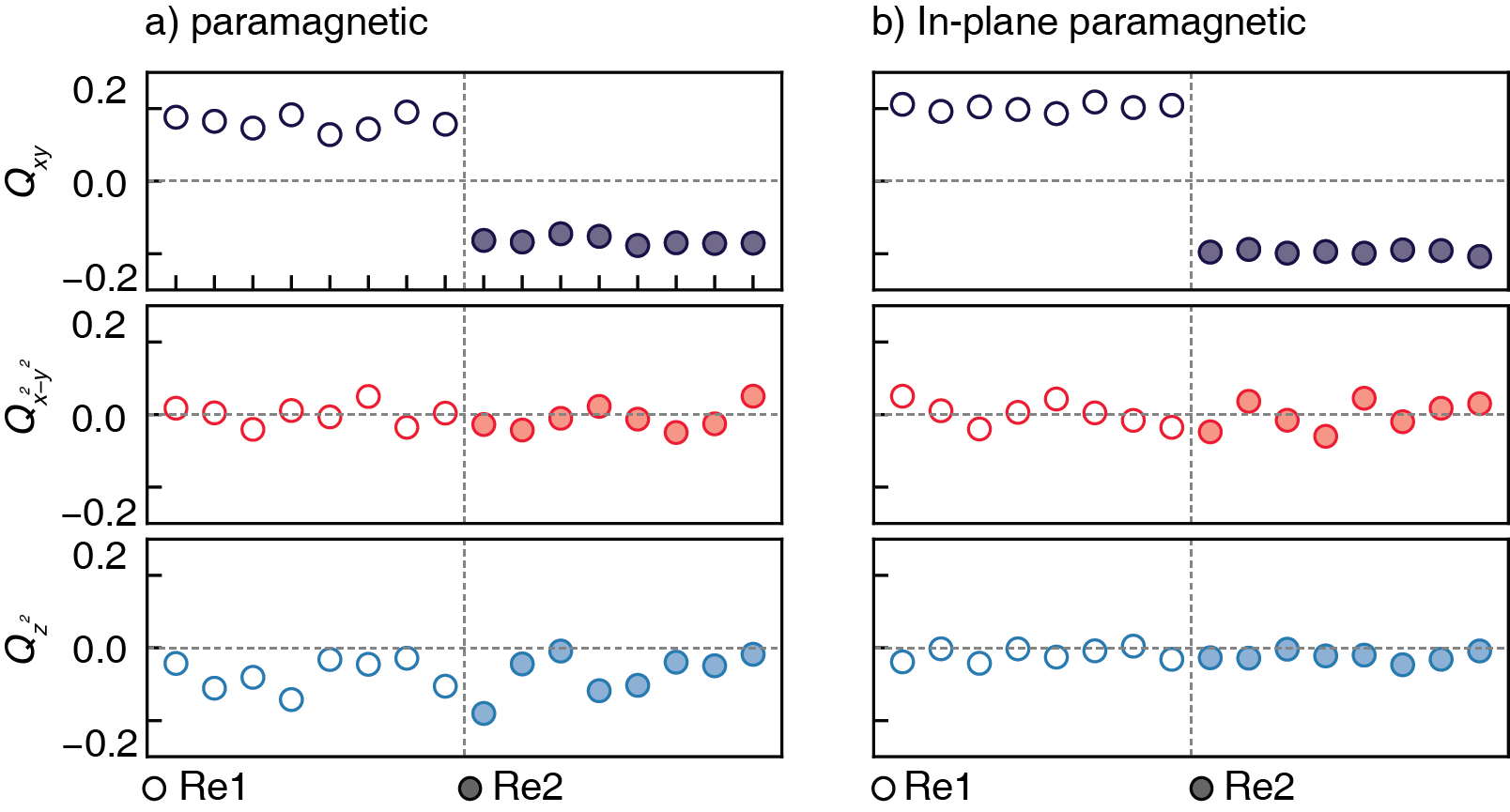}
\centering
\caption{a) $Q_{xy}$, $Q_{x^2-y^2}$, and $Q_{z^2}$ components of the charge quadrupoles in the unit of electron charges for a) fully paramagnetic, and b) in-plane paramagnetic cases. The unfilled data points refer to the Re1 site and the filled points refer to the Re2 site.} 
\label{fig5}
\end{figure*}

Next, we calculate the $Q_{xy}$, $Q_{x^2-y^2}$, and $Q_{z^2}$ charge quadrupole components in the fully and the in-plane disordered paramagnetic states. Our results are plotted in Fig.\,\ref{fig5} with the values shown for all 16 Re sites. The first 8 Re sites in Fig.\,\ref{fig5} correspond to the Re1 site, shown by unfilled data points, while the next 8 Re sites, indicated by the filled data points, correspond to the Re2 site. A ferroic or an anti-ferroic ordering occurs when all the unfilled and filled data points are non-zero and have either the same or opposite sign to each other, respectively. Data points in Fig.\,\ref{fig5} represent an average of 5 supercells with different random configurations of spins. For both the paramagnetic and the in-plane paramagnetic cases the atomic positions are constrained to those obtained in the fully relaxed canted-AFM calculation with the $P4_2/mnm$ space group.

As can be seen from Fig.\,\ref{fig5}a, even in the case of disordered local magnetic dipole moments, the $Q_{xy}$ components order anti-ferroically similar to those of the magnetically ordered unit cells, shown in Fig.\,\ref{fig3}. $Q_{z^2}$ also shows a tendency to partial ferroic order. Conspicuously, the $Q_{x^2-y^2}$ components, while non-zero, do not order in the absence of spin ordering. It can therefore be inferred that, while the $Q_{x^2-y^2}$ components become non-zero at the structural phase transition, they do not order until lower temperatures when the magnetic dipole ordering occurs. This is consistent with our earlier observation that $Q_{x^2-y^2}$, in contrast to $Q_{xy}$ and $Q_{z^2}$, depends strongly on the canting angle. For the hypothetical in-plane paramagnetic phase, interestingly, we find (Fig.\,\ref{fig5}b) full anti-ferroic ordering of the $Q_{xy}$ component. The $Q_{x^2-y^2}$ and $Q_{z^2}$ components are almost zero and can be considered negligible despite the fact that they order in the magnetically ordered unit cell.

\section{Concluding remarks}

In summary, our PBE$+U$ DFT calculations reproduced the experimentally reported low-temperature crystal ($P4_2/mnm$) and magnetic (in-plane canted-AFM) structure of Ba$_2$MgReO$_6$, as well as the previously proposed antiferroic and ferroic ordering of the $Q_{xy}$ and $Q_{z^2}$ quadrupoles on the Re ions. In addition, we found a previously unreported ferroic ordering of the Re $Q_{x^2-y^2}$ quadrupoles. We showed that the Re quadrupoles can be non-zero if either a structural distortion of the local coordination octahedra or the orientation of the magnetic dipole moment lowers the local symmetry from that found in the parent $Fm\bar{3}m$ structure and that their sizes are sensitive to the canting angle of the magnetic dipoles and/or the strength of the spin-orbit coupling. Since the structural distortion and quadrupolar ordering occur experimentally at a higher temperature than the magnetic dipolar ordering, we applied the DFT supercell approach that was recently developed to describe local-moment paramagnetic insulators and found that it successfully captures the SOC-driven insulating behavior even in the absence of magnetic ordering. We used the supercell approach to study the quadrupoles in the low-symmetry structure, both with the local magnetic moments fully disordered and also with them constrained to lie in the easy plane but with random orientation within the plane. In both cases, we found that the ferroic ordering of the $Q_{xy}$ quadrupoles was largely unaffected by the magnetic moment disorder, while the other quadrupolar orders, although still non-zero, were suppressed.  The question of whether the structural distortion or the quadrupole formation provides the primary order parameter at $T_Q$, or indeed if they can be distinguished, remains open since our calculations with both magnetic moment disorder and local structural disorder, which could shed light on this question, did not yield the experimentally relevant insulating state. 

Our results suggest that a study using magnetic resonant elastic x-ray scattering to study the build-up of the in-plane magnetic fluctuations between $T_m$ and $T_Q$, which would manifest as diffuse magnetic peaks, would be fruitful. In addition, measurement of the full spin-wave spectrum of the Re magnetic moments in the magnetically ordered phase below $T_m$ using inelastic neutron scattering would provide valuable insight into the influence of the local magnetic anisotropy and the strength of the magnetic exchange interactions. Finally, we note that the methodology presented here, in which the complex interactions between spin, orbital, and structural degrees of freedom are isolated by separately simulating phases with different types of orderings, provides a general platform for understanding hidden orders and their relationships to the crystal structure and magnetism of materials.

\section{Acknowledgments}

We thank Henrik M. Rønnow, Ivica \v{Z}ivkovi\'{c}, Jana P\'{a}sztorov\'{a}, Rui Soh Jian, and Bruce Normand for many fruitful discussions and comments on the research and the manuscript. This work was funded by the European Research Council (ERC) under the European Union’s Horizon 2020 research and innovation program project HERO grant (No. 810451). Calculations were performed at the Swiss National Supercomputing Centre (CSCS) under project IDs s889 and eth3 and on the EULER cluster of ETH Zürich.

\bibliography{BMRO_2,Nicola}

\end{document}